\documentclass{lmcs}
\usepackage{mik-makro,amsmath,graphicx,forest}
\begin{document}
	
	\newcommand{\blank}{a}
	\newcommand{\ptrees}[1]{\trees_{#1}}
	\newcommand{\trees}{\mathsf{trees}}
		\newcommand{\arity}{\mathsf{arity}}
	\newcommand{\deriv}[1]{\mathsf{deriv}#1}	

\title{It is undecidable if two regular tree languages can be separated by a deterministic tree-walking automaton}
\author{Miko{\l}aj Boja\'nczyk}
\maketitle
\begin{abstract}
	The following problem is shown undecidable: given  regular languages $L,K$ of finite trees, decide if there exists a deterministic tree-walking automaton which accepts all trees in $L$  and rejects all trees in $K$. The proof uses a technique of Kopczy\'nski from~\cite{kopczynski}.
\end{abstract}

\section{Introduction}
Regular languages have  a sort of anti-Rice theorem:  for every natural property $X$, one can decide which  regular languages have property $X$. Examples of such properties include: empty, infinite, universal, commutative, upward closed in the Higman ordering, definable in first-order logic, etc. There are  properties  for which no algorithm is known, e.g.~definable in level $\Sigma_5$ of the first-order quantifier hierarchy  (see~\cite{placesiglog} for a discussion on how algorithms were provided for the first 4 levels), but many believe that with sufficient effort the algorithm will be found. Trees -- at least finite ones -- look similar, with algorithms for properties like emptiness, finiteness, or upward closure being quite straightforward. Of course  trees are always a bit  more challenging, so some questions remain open, e.g.~it is not known if one can decide  which regular tree languages are definable in first-order logic~\cite{DBLP:conf/caap/Thomas84}. Nevertheless,  the prevailing opinion seems to be that the final answer to this and other questions will be ``decidable''.

This paper gives an example of an undecidable property of regular  tree languages, namely this:
	\begin{thm}\label{thm:undecidable-dtwa}
		The following problem is undecidable:
	\begin{itemize}
		\item {\bf Input.} Two regular tree languages, given as bottom-up automata;
		\item {\bf Question.} Can they be separated by a   deterministic tree walking automaton, i.e.~is there a deterministic tree walking automaton which accepts all trees in the first language, and rejects all trees in the second language?
	\end{itemize}
	\end{thm}
The undecidable question in the above theorem is a property not of one, but of two regular tree languages. Questions about separation, like the one above, are currently an important theme in the theory of regular languages, see e.g.~the references in Section 5 of the survey~\cite{placesiglog}.

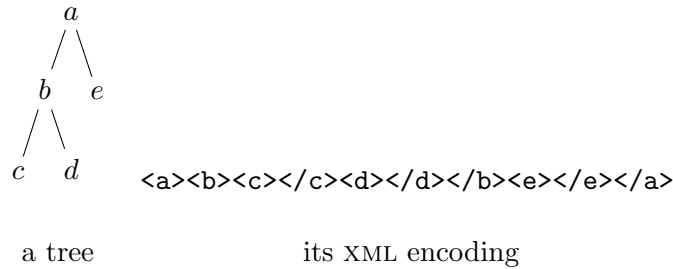
\begin{figure}[htbp]
	\centering
	\begin{tabular}{cc}
	\begin{forest}
		[$a$[$b$[$c$][$d$]][$e$]]   
	\end{forest}
	&  	{\tt<a><b><c></c><d></d></b><e></e></a>} \vspace{0.5cm}\\
		a tree & its {\sc xml} encoding
	\end{tabular}
	\caption{{\sc xml} encoding}
	\label{fig:xml}
\end{figure}

This paper is  closely based on a result by Kopczy\'nski~\cite{kopczynski}, which showed that it is undecidable if two visibly pushdown word languages can be separated by a regular word language. Since a visibly pushdown word language can be viewed as a tree language, Kopczy\'nski's result can be rephrased as follows: it is undecidable if two given two regular tree languages can be separated by a regular property of their {\sc xml} encodings, see Figure~\ref{fig:xml}. Because of the similarity of visibly pushdown languages to pushdown languages, the revolutionary character of  Kopczy\'nski's result was less apparent -- after all,  so many questions about pushdown automata are undecidable (like universality, or more close to this topic, separation by regular languages).

This paper differs only very slightly from~\cite{kopczynski}. Our problem has the same instances (pairs of regular tree languages, also known as visibly pushdown languages), it asks a very similar separation question, and we use the same reduction to prove undecidability. Since our separating mechanism is stronger than the one used by Kopczy\'nski (deterministic tree-walking automata, as opposed to regular properties of the {\sc xml} encoding), we need a stronger lemma to prove correctness of the reduction, but this stronger lemma is simply taken from the literature; thus making the proof slightly shorter than~\cite{kopczynski} but not self-contained.

I would like to thank Sylvain Schmitz for helpful comments on a first draft. 
%

\section{Trees and their automata}	
This section defines basic tree terminology, and introduces the two models of tree automata that will be considered: the stronger model of deterministic bottom-up tree automata, and the weaker model of deterministic tree-walking automata. 
\subsubsection*{Trees and terms.} In this paper, a \emph{ranked alphabet} is a finite set where each element has an associated  arity (a natural number, with zero being used for letters that are used to label leaves).
For a finite ranked aphabet $\Sigma$, define a tree over $\Sigma$ to be a finite, sibling-ordered tree, where every node has a label from $\Sigma$ and  the number of children is the arity of the label.  For $n \ge 0$, define an $n$-ary term over $\Sigma$ to be a tree over the alphabet $\Sigma \cup \set{\text{*}}$, where * is a letter of arity zero that appears exactly $n$ times. Every occurrence of * is called a \emph{port}, the idea is that trees or terms can be substituted into a port.  We write $\trees_n \Sigma$ for the set of $n$-ary terms. In the case $n=0$  of trees we omit the subscript $0$. If $t$ is an $n$-ary term, and $t_1,\ldots,t_n$ are terms, then we write $t(t_1,\ldots,t_n)$ for the term (whose arity is the sum of arities of the terms $t_1,\ldots,t_n$)) obtained from $t$ by substituting $t_i$ for the $i$-th port in $t$. Note that our notion of term uses each argument once, as opposed to the more typical notion which allows each argument to be used several times.

We consider two automaton models for  trees, as described below.
	
	\subsubsection*{Deterministic bottom-up tree automata.}
	A \emph{deterministic bottom-up tree automaton} consists of:  an input ranked alphabet $\Sigma$, a state space $Q$, a set $F \subseteq Q$ of accepting states, and for each letter $a \in \Sigma$  of arity $n$ a transition function:
	\begin{align*}
\delta_a : Q^n \to Q.
	\end{align*}
The automaton is evaluated on a tree in a bottom up way. The state in a tree is obtained by reading the root label, and applying its transition function to the states in the child subtrees.  The language recognised by such an automaton is the set of all trees which are evaluated to  an accepting state. A tree language is called \emph{regular} if it is recognised by such an automaton. 

	\subsubsection*{Deterministic tree-walking automata.} A computation of a deterministic bottom-up tree automaton, as described above, can be viewed as a branching computation, since the state in a node depends on the states in all of its children. In contrast, a tree-walking automaton, as described below, is a sequential device, where a computation has a linear structure. The syntax of a deterministic tree-walking automaton  consists of a ranked input alphabet $\Sigma$, a set of states $Q$, an initial state $q_0 \in Q$, and for each letter $a \in \Sigma$  of arity $n$ a transition function
	\begin{align*}
		\delta_a : \underbrace{Q \times \set{\text{root},1,\ldots,\text{maxarity}}}_{\text{what the automaton sees}}  \to \underbrace{\set{\text{accept,reject}} \cup \big(Q \times \set{\text{parent}, 1,\ldots,n}\big)}_{\text{what the automaton does}},
	\end{align*}
	where maxarity stands for the maximal arity of letters in the input alphabet.
In a given input tree, a \emph{configuration} of the automaton is a pair of the form (state of the automaton, node of the tree). The automaton begins in the configuration which consists of the initial state and the root of the input tree. When in a configuration $(q,v)$, the automaton applies the transition function corresponding to $v$'s label, with the argument to the function being the state $q$ and the child number of $v$ (i.e.~the number $i$ such that $v$ is the $i$-th child of its parent, or ``root'' if $v$ has no parent). Based on the result of the transition function, the automaton chooses to accept/reject the tree, or change its state and make a move to some neighbouring node (or no move at all). In principle, there can be runs that do not accept because the automaton enters a loop, or runs where the automaton walks out of the tree by e.g.~moving to the parent in the root node. As shown in~\cite{DBLP:journals/ipl/MuschollSS06}, every deterministic tree-walking automaton can be converted in polynomial time into one which always ends up by using an accept or reject command.

\section{Undecidability of separation}	
We say that two sets are \emph{separated} by a set $M$ if $M$ contains the first set and is disjoint with the second. The contribution of this paper is the following theorem.
	
The proof of the above theorem uses   a technique  from~\cite{kopczynski}, which shows undecidability for separation of visibly pushdown languages by regular word languages.  As in~\cite{kopczynski}, we reduce from the  following undecidability result, which was shown even under the assumption that the input grammars are deterministic, see Theorem 4.6 in~\cite{DBLP:journals/siamcomp/SzymanskiW76}.
	\begin{thm}\label{thm:undecidable-cfg}
		The following problem is undecidable:
	\begin{itemize}
		\item {\bf Input.} Two context-free word languages, given by grammars.
		\item {\bf Question.} Can they be separated by some regular word language?
	\end{itemize}
	\end{thm}
The reduction we use is actually the same transformation from context-free grammars to tree languages as  used  by Kopczy\'nski in~\cite{kopczynski}, only the  correctness proof is different, since we reduce to a slightly different problem (the problem used by Kopczy\'nski had the same instances, but a weaker class of separating languages, and therefore fewer ``yes'' instances). 

The main result about deterministic tree-walking automata that is needed for the correctness proof is the following lemma on deterministic tree-walking automata, which is taken from~\cite{DBLP:journals/tcs/BojanczykC06}.
\newcommand{\ltrees}{\mathsf{ltrees}}
For a tree language $L \subseteq \trees \Sigma$ define two terms $t,t' \in \trees_n \Sigma$ to be $L$-equivalent if 
\begin{align*}
	s(t(s_1,\ldots,s_n)) \in L \quad \mbox{iff} \quad 	s(t'(s_1,\ldots,s_n)) \in L 
\end{align*}
holds for every $s \in \trees_1 \Sigma$ and $s_1,\ldots,s_n \in \trees \Sigma$. The following Lemma was proved
\footnote{The careful reader will note that~\cite{DBLP:journals/tcs/BojanczykC06} proves a weaker result, namely Lemma 18, which uses a very slightly coarser notion of $L$-equivalence, call it \emph{weak $L$-equivalence}, see page 4 in~\cite{DBLP:journals/tcs/BojanczykC06}. In weak $L$-equivalence, we require that \begin{align*}
	s(t(s_1,\ldots,s_n)) \in L \quad \mbox{iff} \quad 	s(t'(s_1,\ldots,s_n)) \in L 
\end{align*}
holds for every $s \in \trees_1 \Sigma$ and $s_1,\ldots,s_n \in \trees \Sigma$ which satisfy the additional condition that each port is a left child in $s$ and each $s_i$ has at least two nodes. In the proof of Lemma 18, the term $t$ has the property that it is weakly $L$-equivalent to 
\begin{align*}
	\begin{forest}
		[$s$[$t$[$s$[$*$]][$s$[$*$]]]]
	\end{forest}
\end{align*}
 for some $s$ where the only leaf port is a left child. For such terms, weak $L$-equivalence coincides with $L$-equivalence as used in the Rotation Lemma.
}
 in from~\cite{DBLP:journals/tcs/BojanczykC06}.

\begin{lem}[Rotation Lemma]  Let $\Sigma$ be a ranked alphabet, which contains a letter $a$ of rank 2 and a letter $c$ of rank 0. Let $L$ be a tree language over $\Sigma$ which is recognised by a deterministic tree-walking automaton. There exists some $t \in \trees_2\set{a,c}$  such that following two terms are $L$-equivalent:
	
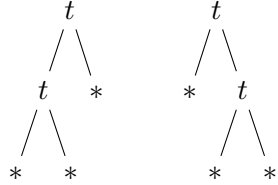
\begin{align*}
		\begin{forest} 
		[$t$[$t$[$*$] [$*$]][$*$]]
		\end{forest}
\qquad
		\begin{forest} 
		[$t$ [$*$] [$t$[$*$] [$*$]]]
		\end{forest}
	\end{align*}
\end{lem}


\newcommand{\kopc}[1]{\mathsf{kop}(#1)}
\subsubsection*{Kopczy\'nski obfuscation.} We now present the reduction from separation of context-free word languages by a regular word language (the problem in Theorem~\ref{thm:undecidable-cfg}) to separation of regular tree languages by a deterministic tree-walking automaton. 
	Consider a context-free grammar $G$ in Chomsky normal form, with terminals $\Gamma$ and nonterminals $N$. Since we use Chomsky normal form, nonterminals get transformed into pairs of nonterminals, and therefore we can view $\Gamma$ as ranked letters of arity zero, and $N$ as ranked letters of arity $2$, and we  can view derivations of the grammar as trees in $\trees(\Gamma \cup N)$.
	
Choose some fresh letters $a,c$, of arities 2 and 0 respectively.  	The \emph{Kopczy\'nski obfuscation} of $G$, denoted by $\kopc G$, is the set of all trees that can be obtained from some derivation of the grammar, and replacing each nonterminal by a binary term over the alphabet $\set{a,c}$, possibly using different terms for different occurrences of nonterminals. A more formal definition is that
\begin{align*}
	\kopc G = \bigcup_{\text{$t$ a derivation of $G$}} \kopc t,
\end{align*}
while $\kopc t$ is the set of trees over alphabet $\Gamma \cup \set{a,c}$ defined by
	\begin{align*}
		\begin{array}{llll}
			\kopc \sigma  & = & \set\sigma\\
			\kopc {\sigma(t_1,t_2)} & = & \set{s(s_1,s_2) : s \in \trees_2\set{a,c}, s_1 \in \kopc {t_1}, s_2 \in \kopc{t_2}} 
		\end{array}
	\end{align*}
	where the first line is used for trees with just one node, and the second line for other trees. We use the name Kopczy\'nski because mapping a grammar to its Kopczy\'nski obfuscation was the reduction used in~\cite{kopczynski}, as it is also in this paper.	It is not difficult to see that the obfuscation is a regular tree language and that a tree automaton for the obfuscation can be computed based on the grammar. The following lemma shows that taking the Kopczy\'nski obfuscation reduces the undecidable problem in Theorem~\ref{thm:undecidable-cfg} to the problem in Theorem~\ref{thm:undecidable-dtwa}, thus proving undecidability of the latter.
\begin{lem}\label{lem:reduction}
	Let $G,H$ be context free grammars, with terminals $\Gamma$. 	The following conditions are equivalent:
	\begin{enumerate}
		\item 	The tree languages 
	\begin{align*}
		\kopc G, \kopc H \subseteq \trees (\Gamma \cup \set{a,c})
	\end{align*}
	 can be separated by a deterministic tree-walking automaton.
	 \item  The word languages  
	 \begin{align*}
	 	L(G),L(H) \subseteq \Gamma^*
	 \end{align*}
	 generated by these grammars can be separated by a regular word language.
	\end{enumerate}
\end{lem}

The implication from 2 to 1 in the above lemma is straightforward. This is because for every regular word language $L$, in particular the separator, there is  a deterministic tree-walking automaton that accepts an input tree if and only if $L$ contains the sequence of leaves  read from left to right. The idea is to use depth-first search, see e.g.~Example 1 in~\cite{DBLP:conf/lata/Bojanczyk08}.

It remains to prove the converse implication from 1 to 2. Here our task is more difficult than in~\cite{kopczynski}, because deterministic tree-walking automata are relatively powerful, and can be quite challenging to prove that they cannot do something.
We use the following corollary of the Rotation Lemma. For $t \in \trees_2 \Sigma$,  define $t^*$ to be the smallest set of terms that contains $*$ (a unary term with the port in the root) and  which  is closed under composition with $t$ in the following sense:
\begin{align*}
	t_1,t_2 \in t^* \qquad \mbox{implies} \qquad t(t_1,t_2) \in t^*.
\end{align*}

		\begin{lem}\label{lem:rot-cor}
			Let $L \subseteq \trees \Sigma$ and $t$ be as in the Rotation Lemma and let $\Gamma$ be the rank 0  symbols in $\Sigma$. There is a regular word language $K \subseteq \Gamma^*$ such that
	\begin{align*}
		a_1 \cdots a_n \in K \qquad \mbox{iff} \qquad s(a_1,\ldots,a_n) \in L
	\end{align*}
holds	 for every $n \ge 2$,  $a_1,\ldots,a_n \in \Gamma$ and $n$-ary $s \in t^*$.
		\end{lem}
		
Before proving the above lemma, note that it implies that as long as $s$ is taken from $t^*$, then membership of $s(a_1,\ldots,a_n)$  in $L$ does not depend on the branching structure of $s$, but only on the number of ports.

\begin{proof}
	For $a_1,\ldots,a_n \in \Gamma$, define $\mathsf{comb}(a_1,\ldots,a_n)$ to be the following tree:
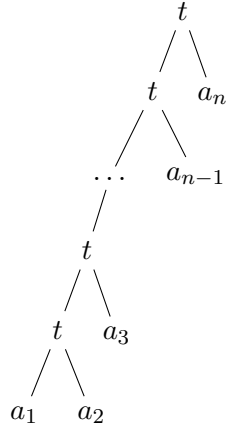
\begin{center}
	\begin{forest}
		[$t$[$t$[$\ldots$[$t$[$t$[$a_1$][$a_2$]][$a_3$]][,phantom]][$a_{n-1}$]][$a_n$]]
	\end{forest}
\end{center}
Every two binary trees with the same number of leaves can be transformed into each other via a sequence of rotations. Therefore, repeated application of  the Rotation Lemma shows that every $n$-ary $s \in t^*$ satisfies
\begin{align*}
	s(a_1,\ldots,a_n)  \in L \qquad \mbox{iff} \qquad \mathsf{comb}(a_1,\ldots,a_n) \in L.
\end{align*}
To complete the proof, it suffices to show that 
\begin{align*}
 K =	\set{a_1 \cdots a_n \in \Gamma^*: \mathsf{comb}(a_1,\ldots,a_n) \in L } 
\end{align*}
is a regular word language. Since  deterministic tree-walking automata can only recognise regular tree languages, see e.g.~Fact 1 in~\cite{DBLP:conf/lata/Bojanczyk08}, there is a bottom-up tree automaton $\Aa$  that recognises $L$. We define a deterministic word automaton recognising $K$ as follows. The states are the same as in $\Aa$ plus a special initial state.  When the automaton is in the initial state and reads a letter $\sigma \in \Gamma$, it moves to the state of $\Aa$ after reading a one node tree $\sigma$. When the automaton is in a state of $\Aa$, then the  transition function is defined by
\begin{align*}
	\delta(q,\sigma) = t(q,\sigma) \qquad \mbox{for $\sigma \in \Gamma$}
\end{align*}
where $t(q,\sigma)$ is the state of $\Aa$ after reading a tree  obtained from  $t(*,\sigma)$ by putting some tree evaluated to $q$ into the port. By definition, this word automaton maps a word $a_1 \cdots a_n \in \Gamma^*$ to the state of the tree automaton $\Aa$ after reading the tree $\mathsf{comb}(a_1,\ldots,a_n)$, and therefore the language $K$ is regular.
\end{proof}

Using the above lemma, we complete the implication from 1 to 2 in Lemma~\ref{lem:reduction}.
	Suppose that $\kopc G$ can be separated from $\kopc H$ by some deterministic tree-walking automaton recognising a language   $L \subseteq \trees(\Gamma \cup \set{a,c})$. Apply the Rotation Lemma to $L$, yielding $t$, and apply Lemma~\ref{lem:rot-cor} yielding a regular word language $K \subseteq \Gamma^*$. We claim that  $K$ separates the context-free word languages generated by $G$ and $H$. Indeed, suppose that  $a_1 \cdots a_n$ is generated by $G$. By taking the corresponding derivation and replacing each nonterminal by $t$, we see that  there is some $n$-ary term $s \in t^*$ such that
	\begin{align*}
		s(a_1,\ldots,a_n) \in \kopc G.
	\end{align*}
	Since $\kopc G$ is contained in $L$, it follows that  $a_1 \cdots a_n \in K$. Conversely, if $a_1 \cdots a_n$ is generated by $H$, then there is some  $n$-ary term $s \in t^*$ such that
	\begin{align*}
		s(a_1,\ldots,a_n) \in \kopc H.
	\end{align*}
	Since $\kopc H$ is disjoint with $L$, it follows that $a_1 \cdots a_n \not \in K$.
This completes the proof of Lemma~\ref{lem:reduction}, and therefore also of Theorem~\ref{thm:undecidable-dtwa}.

\section{What is the scope of the technique?}
The  proof of Theorem~\ref{thm:undecidable-dtwa} works not just for deterministic tree-walking automata, but also for any class of regular languages $\Ll$ that satisfies the Rotation Lemma and is strong enough to express properties like: ``the sequence of leaves, when read from left to right, belongs to a regular language $K$''. 
 However, this makes the technique sound more powerful than it is: the Rotation Lemma is a very strong lemma, and seems to hold only for deterministic tree-walking automata and their special cases. For example, the Rotation Lemma fails for nondeterministic tree-walking automata, and all fragments of first-order logic beyond Boolean combinations of $\Sigma_1$ sentences, for which separation is decidable~\cite{goubault}.

 It seems therefore that the technique of Kopczy\'nski obfuscation is exhausted by deterministic tree-walking automata. As an example, we claim that one can find:
 \begin{itemize}
 	\item a grammar $G$ generating the palindromes; and
 	\item a grammar $H$ generating  the non-palindromes;
 \end{itemize}
 such that the Kopczy\'nski obfuscations $\kopc G$ and $\kopc H$ can be separated by a nondeterministic tree-walking automaton, thus showing that the reduction in Lemma~\ref{lem:reduction} fails for nondeterministic tree-walking automata. The trick is to choose the grammars so that their derivations have shapes  as in Figure~\ref{fig:derivations}; then the technique from  Lemma 2 in~\cite{DBLP:journals/tcs/BojanczykC06} can be used to separate $\kopc G$ from $\kopc H$. This  counterexample also works for other separators, e.g.~for first-order logic. The counterexample only means that the same reduction cannot be used, but the problem might still  be undecidable.
 
 \begin{figure}[htbp]
 	\centering
	\begin{tabular}{cc}
	\begin{forest}
		[$.$[$.$[$.$][$.$[$.$[$.$][$.$[$.$][$.$]]][$.$]]][$.$]]
	\end{forest} &
	\begin{forest}
		[$.$[$.$][$.$[$.$[$.$][$.$[$.$[$.$][$.$]][$.$]]][$.$]]]
	\end{forest}\\
		derivations in $G$& derivations in $H$
	\end{tabular}
 	\caption{In a derivation from $G$, the right child of the root is a leaf, while in a derivation from $H$, the left child of the root is a leaf.}
 	\label{fig:derivations}
 \end{figure}
 
 \subsubsection*{Conclusion.} The conclusion is that some questions about regular tree languages can indeed be undecidable. The particular undecidability proof in this paper strongly depends on the Rotation Lemma -- which is true essentially only for deterministic tree-walking automata -- and on separation.  To highlight the role of separation, consider the class  $\Ll$ of regular tree languages $L$ such that $t \in L$ depends only on  the sequence of leaves in $t$, read from left to right. Then membership of regular tree language in $\Ll$ is decidable (see Theorem 1 in~\cite{DBLP:journals/tcs/Wilke96} for a stronger result) but separation of two regular tree languages by $\Ll$ is undecidable, using the same proof as here or in~\cite{kopczynski}.

\bibliographystyle{alpha}
\bibliography{bib}

\end{document}